\documentclass[preprint,preprintnumbers,amsmath,amssymb]{revtex4}
\usepackage{graphicx}

\begin{document}

\title{Energy transfer efficiency in the FMO complex strongly coupled to a vibronic mode}

\author{ Lev G. Mourokh$^{1}$ and Franco Nori$^{2,3}$}

\affiliation{ $^1$ Department of Physics, Queens College, The City
University of New York, Flushing, New York 11367, USA \\ $^2$ CEMS,
RIKEN, Saitama, 351-0198, Japan \\
$^3$ Physics Department, The University of Michigan, Ann Arbor, MI 48109-1040, USA \\
}

\date{\today}

\begin{abstract}

Using methods of condensed matter and statistical physics, we
examine the transport of excitons through the Fenna-Matthews-Olson
(FMO) complex from a receiving antenna to a reaction center. Writing
the equations of motion for the exciton creation/annihilation
operators, we are able to describe the exciton dynamics, even in the
regime when the reorganization energy is of the order of the
intra-system couplings. In particular, we obtain the well-known
quantum oscillations of the site populations. We determine the
exciton transfer efficiency in the presence of a quenching field and
protein environment. While the majority of the protein vibronic
modes are treated as a heat bath, we address the situation when
specific modes are strongly coupled to excitons and examine the
effects of these modes on the quantum oscillations and the energy
transfer efficiency. We find that, for the vibronic frequencies
below 16 meV, the exciton transfer is drastically suppressed. We
attribute this effect to the formation of ``polaronic states'' where
the exciton is transferred back and forth between the two pigments
with the absorption/emission of the vibronic quanta, instead of
proceeding to the reaction center. The same effect suppresses the
quantum beating at the vibronic frequency of 25 meV. We also show
that the efficiency of the energy transfer can be enhanced when the
vibronic mode strongly couples to the third pigment only, instead of
coupling to the entire system.

\end{abstract}

\maketitle

\section{Introduction}

The effective energy transfer in photosynthetic complexes has been
one of the focal points of experimental and theoretical studies
during recent years \cite{NatChemRev,FrancoRev}. The light energy is
absorbed by the pigments in the antenna systems and subsequently
transferred to a reaction center, where the created electron-hole
pairs are separated and the energy is converted to chemical
compounds. The energy conversion efficiency of this process can
reach 99\% \cite{Excitons,Blank}. The chromophore complexes located
between the antenna and the reaction center are of special interest,
because of the recent observation of the long-lived quantum
coherence in the Fenna-Matthews-Olson (FMO) complex of green sulfur
bacteria \cite{Engel1,Engel2} and marine cryptophyte algae
\cite{Scholes1,Scholes2}.

The FMO complex \cite{FMO} is one of the most studied
photosynthesis-related systems. It is a trimer with each unit
consisting of seven bacteriochlorophyll a (BChl-a) molecules and
with three more BChls between the units (so-called the eighth
BChls). This molecular system is placed between an antenna and the
reaction center where the charges are separated. The light harvested
in the antenna excites one of the BChls and this excitation
propagates through the complex until it reaches the reaction center.
This energy transfer has been addressed in numerous theoretical
works
\cite{Jang1,Renger,Ishizaki1,Mukamel,Cao1,Ishizaki2,Sarovar1,Olaya1,Cao2,Nalbach1,Jang2,Nalbach2,Nalbach3,Roden1,Olbrich1,Olbrich2,Nazir1,Nazir2,Alicki}
employing various types of approximations. BChls are coupled to the
protein environment and its role in the excitation transfer has been
widely discussed. In particular, it was suggested that the
interaction with the environment can assist exciton transport
\cite{Plenio1,Aspuru1,Aspuru2,Roden2,Olaya2,Olaya3} and, moreover,
can lead to a new type of exitonic-vibrational coherence
\cite{Mancal1,Mancal2,Plenio2,Tiwari1}. However, it is still open
question how this type of coherence affects the efficiency of the
exciton transfer through the chromophore network.

In this paper, we examine the exciton propagation through the system
of many inter-coupled chromophores to the reaction center in the
presence of an external excitation, radiation heat bath, quenching
field, and protein environment considered as a system of independent
oscillators. The main focus here is to the coupling to a specific
vibronic mode which cannot be treated as a heat bath. The coupling
strengths of all the surrounding components are very different, so
various levels of approximations should be used. In particular, the
contributions of the quenching and radiation fields, as well as that
of the reaction center, are calculated perturbatively within the
secular approximation. The reorganization energy of the protein
environment is of the order of the inter-chromophore couplings, so
we have to incorporate non-Marcovian effects. However, in the case
of slow protein motion and high enough temperatures, the dynamics
under the time integral can be simplified. We do not apply this
approximation for the strongly-coupled vibronic mode and also take
into account the interplay of the last two processes. For all the
above interactions, we take into account the contributions of both
diagonal and off-diagonal elements of the density matrix.

The couplings of the system to the external light source and to a
reaction center are explicitly included in our Hamiltonian.
Therefore, we are able to calculate the rate of energy transfer to
the reaction center and the rate of the energy absorbtion by the
system and, correspondingly, to {\it directly} determine the
efficiency of the energy transfer. Previously,
\cite{Jesenko1,Witt1}, the efficiency was calculated indirectly by
the population of the last chromophore in the chain. It should be
also emphasized that our approach is not restricted to the
single-exciton-propagation case, and the chromophore network can
contain as many excitons as the number of chromophores. Equations
obtained for the general case are applied to the FMO complex. We
show that for certain frequencies of the vibronic mode, the energy
transfer is strongly suppressed. To explain these results, we
determine the time dependencies of the chromophore populations and
show that, for these frequencies, the excitation does not
effectively reach the chromophore coupled to the reaction center,
despite the fact that it has the lowest energy. We argue that this
effect is the result of the exitonic-vibrational coherence
\cite{Mancal1,Mancal2,Plenio2,Tiwari1}, when the {\it polaronic
mode} is formed and the excitation is transferred between two
excitonic states with the emission/absorption of vibronic quanta,
instead of proceeding to the reaction center. Eventually, the
excitation energy is dissipated. The same effect can suppress the
quantum beating which occurs in the populations of the first and
second pigments. At a certain frequency of the vibronic mode, the
polaronic mode forms between the first and sixth pigments and the
exciton is transferred to the sixth chromofore instead of the second
one. However, if the vibronic mode is strongly coupled to a specific
(third) pigment, the efficiency of the energy transfer is even
enhanced.

The rest of the paper is structured as follows. Section II contains
the Hamiltonian of the system. In Section III, we determine the
density matrix and derive its equations of motion. The efficiency of
the energy transfer is defined in Section IV. This approach is
applied to the FMO complex in Section V, where we find the
dependence of the energy transfer efficiency on the frequency of the
vibronic mode strongly coupled to the system and calculate the time
dependencies of the chromophores populations. Section VI contains
the conclusions of our work.

\section{Hamiltonian}

We start from the general description of the exciton transfer
through a system of $N$ chromophores when these are coupled to each
other, and each one of them can be coupled to the light source,
reaction center, quenching and blackbody radiation fields, as well
as the protein environment. In addition, the interaction with a
specific strong vibronic mode is included. The Hamiltonian of this
system consists of the following components:

(i) Unperturbed part
\begin{equation}
H_0 = \sum_n \epsilon_n a_n^\dag a_n + \sum_{m\neq n} V_{mn}
a_m^\dag a_n - \sum_n ( F_n e^{i\omega_0 t} a_n + F_n^*
e^{-i\omega_0 t} a_n^\dag ), \label{H0}
\end{equation}
where $a_n^\dag$ and $a_n$ are the creation and annihilation
operators for the excitations of the $n$-th chromophore, $E_n$ is
the excitation energy, $V_{mn}$ is the inter-chromophores energy
transfer amplitude, and $F_n$ and $\omega_0$ are the coupling
strengths and frequency of the external light, respectively. It
should be noted that the total number of excitations in the system
depends on the coupling to the light source and can be as many as
the number of chromophores. However, each of them can only be
single-populated.

(ii) Coupling to the reaction center
\begin{equation}
H_{{\rm trap}} = - \sum_n\sum_k \left( g_{kn}b_k^\dag a_n +
g^*_{kn}a_n^\dag b_k \right), \label{Htrap}
\end{equation}
where $b_k^\dag$ and $b_k$ are the creation and annihilation
operators for the excitations at the reaction center having its own
Hamiltonian
\begin{equation}
H_{{\rm RC}} = \sum_k \epsilon_kb_k^\dag b_k . \label{HRC}
\end{equation}

(iii) Interaction with the blackbody radiation and quenching fields
\begin{equation}
H_{{\rm Rec}} = - \sum_n \left( Q_na_n^\dag  + Q^*_na_n \right),
\label{HRec}
\end{equation}
where the operator
\begin{equation}
Q_n = d_n ( {\cal E}_{{\rm rad}} + {\cal E}_{{\rm quen}} ), \label
{Qn}
\end{equation}
is proportional to the sum of the fields multiplied by the dipole
moment $d_n$ of the $n$-chromophore. The Hamiltonians of the
radiation heat bath, $H_{{\rm Rad}}$, and the quenching field,
$H_{{\rm quen}}$, describe the free evolution of these degrees of
freedom.

(iv) Coupling to the protein environment
\begin{equation}
H_{{\rm e-ph}} = - \sum_{j,n} C_{jn} m_j\omega_j^2 x_j a_n^\dag a_n
. \label{Hph}
\end{equation}
We describe the environment, having the Hamiltonian
\begin{equation}
H_{{\rm env}} = \sum_j \left( \frac{p_j^2}{2 m_j} +
\frac{m_j\omega_j^2 x_j^2}{2} \right), \label{Henv}
\end{equation}
as a set of independent harmonic oscillators with the position
($x_j$) and momentum ($p_j$) operators. The $j$-th oscillator has a
mass $m_j$ and a frequency $\omega_j$. Here $C_{jn}$ are the
coupling strengths of the $j$-th phonon mode and the exciton at the
$n$-th site.

(v) Coupling to a specific vibronic mode
\begin{equation}
H_{{\rm e-vib}} = - \sum_n C_{n} M\Omega^2 X a_n^\dag a_n  ,
\label{Hvib}
\end{equation}
with the Hamiltonian
\begin{equation}
H_{{\rm vib}} =  \frac{P^2}{2 M} + \frac{M\Omega^2 X^2}{2} ,
\label{Hvib}
\end{equation}
involving the position $X$ and momentum $P$ operators. Here, $M$,
$\Omega$ and $C_{n}$ are the mass, frequency, and the coupling
strengths, respectively, associated with this vibronic mode.

The time dependence of the unperturbed Hamiltonian, Eq.~(\ref{H0}),
can be removed by means of the unitary transformation,
\begin{equation}
U = \exp\left(-i \sum_m N_m \omega_0 t\right) = \prod_m \exp(-i N_m
\omega_0 t),
\end{equation}
where  $N_m = a_m^\dag a_m$. Accordingly, the total Hamiltonian has
the form
\begin{eqnarray}
H = \sum_n (\epsilon_n - \omega_0) a_n^\dag a_n + \sum_{m\neq n}
V_{mn} a_m^\dag a_n - \sum_n ( F_n a_n + F_n^*  a_n^\dag )
\nonumber\\  - \sum_n( e^{i\omega_0 t} Q_n a_n^\dag + e^{-i\omega_0
t} Q_n^\dag a_n ) -
 \sum_{k,n} ( e^{-i\omega_0 t} g_{kn} b_k^\dag a_n +
e^{i\omega_0 t} g_{kn}^* a_n^\dag b_k ) \nonumber\\ - \sum_{j,n}
C_{jn} m_j \omega_j^2 x_j a_n^\dag a_n - \sum_{n} C_{n} M \Omega^2 X
a_n^\dag a_n \nonumber\\ +  H_{{\rm env}} + H_{{\rm vib}} + H_{{\rm
RC}} + H_{{\rm Rad}} + H_{{\rm quen}}. \label{Htot}
\end{eqnarray}

\section{Density matrix and rate equations}

The unperturbed Hamiltonian can be numerically diagonalized with the
determination of its eigenenergies and eigenfunctions, as
\begin{equation}
H_0 |\mu\rangle = E_{\mu} |\mu \rangle. \label{Eig}
\end{equation}
Accordingly, we can construct the density matrix in the form
\begin{equation}
\rho_{\mu\nu} = | \mu \rangle \langle \nu |, \label{rho}
\end{equation}
and express all operators in terms of this matrix. In particular,
the exciton operators are given by
\begin{eqnarray}
a_m = \sum_{\mu\nu} a_{m;\mu\nu} \rho_{\mu\nu} = \sum_{\mu\nu}
\langle \mu|a_m |\nu\rangle \rho_{\mu\nu}, \; N_m = a_m^\dag a_m =
\sum_{\mu\nu} \langle \mu|N_m|\nu\rangle \rho_{\mu\nu}.
\end{eqnarray}
Correspondingly, the total Hamiltonian of the system can be written
as
\begin{equation}
H = H_0 - \sum_{\mu\nu} {\cal A}_{\mu\nu} \rho_{\mu\nu},
\end{equation}
where ${\cal A}_{\mu\nu}$ includes contributions of all terms of
Eq.~(\ref{Htot}) not involved in $H_0$. It should be noted that the
external light source is already included in $H_0$ and the basic
states are determined accordingly.

We treat the density matrix elements $\rho_{\mu\nu}$ as Heisenberg
operators and the corresponding equations of motion are given by
\begin{equation}
i \dot{\rho}_{\mu\nu} = [\rho_{\mu\nu}, H ]_- = -\, \omega_{\mu\nu}
\rho_{\mu\nu} - \sum_{\alpha} ( {\cal A}_{\nu\alpha}
\rho_{\mu\alpha} - {\cal A}_{\alpha \mu} \rho_{\alpha \nu} ),
\label{em}
\end{equation}
where $\omega_{\mu\nu} = E_{\mu} - E_{\nu}.$

To evaluate specific contributions to Eq.~(\ref{em}), we apply the
approach introduced in Ref.~\cite{SmJCP2011} where the set of exact
non-Markovian equations was derived. The coupling strengths of the
chromophores to surrounding fields are different, and,
correspondingly, various levels of approximations can be used. The
details of the calculations are given in the Appendix.

The time evolution of the off-diagonal ($\mu\neq \nu$) elements of
the exciton matrix $\langle \rho_{\mu\nu}\rangle (t)$ is given by
\begin{equation}
\rho_{\mu\nu}(t) = \exp \left[ i\omega_{\mu\nu}\,t\right] \exp
\left[ -(\bar{\lambda}^{{\rm ph}}_{\mu\nu} + \bar{\lambda}^{{\rm
vib}}_{\mu\nu})\,T\,t^2\right] \exp \left[ - \Gamma_{\mu\nu}\,
t\right] \rho_{\mu\nu}(0),
\end{equation}
where $\bar{\lambda}^{{\rm ph}}_{\mu\nu}$ and $\bar{\lambda}^{{\rm
vib}}_{\mu\nu}$ are the reorganization energies associated with the
phonon heat bath and the strongly-coupled vibronic mode,
respectively; and the dephasing rate
\begin{equation}
\Gamma_{\mu\nu} = \Gamma^{{\rm ph}}_{\mu\nu} + \Gamma^{{\rm
vib}}_{\mu\nu} + \Gamma_{\mu\nu}^{{\rm rec}} + \Gamma_{\mu\nu}^{{\rm
trap}},
\end{equation}
includes contributions of all processes involved.

The exciton distribution  $\langle \rho_{\mu}\rangle$  over the
eigenstates $|\mu\rangle $ of the Hamiltonian $H_0$ evolves
according to the equation
\begin{equation}
\langle \dot{\rho}_{\mu}\rangle + \gamma_{\mu}\langle
\rho_{\mu}\rangle = \sum_{\alpha} \gamma_{\mu\alpha} \langle
\rho_{\alpha} \rangle,
\end{equation}
where the  relaxation matrix $\gamma_{\mu\alpha}$ contains
contributions, $\gamma^{{\rm ph}}_{\mu\alpha}$ and $\gamma^{{\rm
vib}}_{\mu\alpha}$, of the non-diagonal environment and vibronic
operators, as well as contributions of recombination,
$\gamma_{\mu\alpha}^{{\rm rec}}$, and trapping,
$\gamma_{\mu\alpha}^{{\rm trap}}$, processes, as
\begin{equation}
\gamma_{\mu\alpha} = \gamma^{{\rm ph}}_{\mu\alpha} + \gamma^{{\rm
vib}}_{\mu\alpha} + \gamma_{\mu\alpha}^{{\rm rec}} +
\gamma_{\mu\alpha}^{{\rm trap}},
\end{equation}
The density relaxation rate is given by
\begin{equation}
\gamma_{\mu} = \sum_{\alpha} \gamma_{\alpha\mu}.
\end{equation}
The steady-state exciton distribution $\rho_{\mu}^0$ can be found
from the equation
\begin{equation}
\gamma_{\mu} \rho_{\mu}^0 = \sum_{\alpha}
\gamma_{\mu\alpha}\rho_{\alpha}^0,
\end{equation}
taking into account the normalization condition $\sum_{\mu}
\rho_{\mu}^0 = 1.$

\section{Energy-transfer efficiency}

We define the energy-transfer efficiency as a ratio of the average
rate of the energy transmission going to the reaction center to the
total rate of electromagnetic energy absorption by the system. The
first quantity is given by
\begin{equation}
{\cal W}_{{\rm RC}} = \frac{d}{dt} E_{{\rm RC}} = \sum_k \epsilon_k
\langle \dot{N}_k \rangle, \label{ERC}
\end{equation}
where $N_k = b_k^\dag b_k$. The interaction of the system with the
monochromatic light source can be rewritten in terms of the electric
field strength, ${\cal E}_n(t) = F_n e^{i\omega_0 t} + H.c.,$ and
the polarization ${\cal P}_n = a_n^\dag + a_n$, as $H_F = - \sum_n
{\cal E}_n(t) {\cal P}_n.$ Thus, the rate of the light energy
absorption has the form
\begin{equation}
{\cal W} = - \sum_n \langle {\cal P}_n \dot{\cal E}_n(t)\rangle
\simeq  - i\omega_0 \sum_n \langle F_n e^{i\omega_0 t} a_n - F_n^*
e^{-i\omega_0 t} a_n^\dag \rangle. \label{Wabs}
\end{equation}
This energy can be determined using the equation of motion for the
operator of the total number of excitations, $\sum_m N_m$, and can
be written in the form of the balance relation:
\begin{equation}
{\cal W} = {\cal W}_m + {\cal W}_k + {\cal W}_{{\rm Rec}} = \omega_0
\sum_m \langle \dot{N}_m \rangle + \omega_0 \sum_k \langle \dot{N}_k
\rangle  +  i \omega_0 \sum_m \langle e^{-i\omega_0 t} Q_m^\dag a_m
- e^{i\omega_0 t} a_m^\dag Q_m \rangle. \label{Wabs2}
\end{equation}
Correspondingly, the energy transfer efficiency is given by
\begin{equation}
\eta = {\cal W}_{{\rm RC}}/{\cal W}. \label{Eff}
\end{equation}

For the steady state, the total number of excitons in the system is
constant, so ${\cal W}_m = 0$. ${\cal W}_{{\rm RC}}$, ${\cal W}_k$,
and ${\cal W}_{{\rm Rec}}$ can be calculated similar to the
relaxation rates of Appendix and they have forms
\begin{eqnarray}
{\cal W}_{{\rm RC}} = \sum_n \sum_{\mu\nu} (\omega_0 -
\omega_{\mu\nu}) |a_{n;\mu\nu}|^2 \times \nonumber \\ \left\{
\Gamma_n^{{\rm trap}}(\omega_0 + \omega_{\mu\nu})\left[ 1 +
n(\omega_0 + \omega_{\mu\nu})\right] \langle \rho^0_{\nu}\rangle -
\Gamma_n^{{\rm trap}}(\omega_0 - \omega_{\mu\nu})n(\omega_0 -
\omega_{\mu\nu}) \langle \rho^0_{\mu}\rangle \right\} , \label{WRC}
\end{eqnarray}
\begin{eqnarray}
{\cal W}_k = \omega_0 \sum_n \sum_{\mu\nu} |a_{n;\mu\nu}|^2 \times
\nonumber \\ \left\{ \Gamma_n^{{\rm trap}}(\omega_0 +
\omega_{\mu\nu})\left[ 1 + n(\omega_0 + \omega_{\mu\nu})\right]
\langle \rho^0_{\nu}\rangle - \Gamma_n^{{\rm trap}}(\omega_0 -
\omega_{\mu\nu}) n(\omega_0 - \omega_{\mu\nu}) \langle
\rho^0_{\mu}\rangle \right\} , \label{Wk}
\end{eqnarray}
and
\begin{eqnarray}
{\cal W}_{{\rm Rec}} = 2 \omega_0 \sum_n \sum_{\mu\nu}
|a_{n;\mu\nu}|^2 \times \nonumber \\ \left\{ \chi_n''(\omega_0 +
\omega_{\mu\nu})\left[ 1 + n(\omega_0 - \omega_{\mu\nu})\right]
\langle \rho^0_{\nu}\rangle - \chi_n''(\omega_0 - \omega_{\mu\nu})
n(\omega_0 - \omega_{\mu\nu}) \langle \rho^0_{\mu}\rangle \right\} .
\label{WRec}
\end{eqnarray}

\section{FMO complex}

In this Section, we apply the equations obtained above to a specific
system, FMO complex containing seven pigments. We assume that the
external light creates the exciton in the Bchl 1 and the reaction
center is coupled to the Bchl 3. We ignore the eighth Bchl
\cite{8Bchl} in this work because its role in the energy transfer is
not clear yet. In the absence of an external light source, the
energies of the seven exciton sites with respect to the lowest one
and the transfer matrix elements (in meV) are given by
\cite{Adolphs}
\begin{equation}
E_n + V_{mn} =
\begin{pmatrix}
29.76  & -10.87 & 0.68  & -0.73 & 0.83  & -1.7  & -1.23 \\
-10.87 & 39.06  & 3.73  & 1.02  & 0.09  & 1.46  & 0.53  \\
 0.68  & 3.73   & 0     & -6.63 & -0.27 & -1.19 & 0.74  \\
-0.73  & 1.02   & -6.63 & 16.12 & -8.77 & -2.11 & -7.85 \\
0.83   & 0.09   & -0.27 & -8.77 & 35.34 & 10.06 & -0.16 \\
 -1.7  & 1.46   & -1.19 & -2.11 & 10.06 & 53.94 & 4.92  \\
-1.23  & 0.53   & 0.74  & -7.85 & -0.16 & 4.92  & 30.38
\end{pmatrix}
\label{Vmn}
\end{equation}
In our model, each Bchl can be populated, so we have a total of 128
(=$2^7$) basic states. The energies of the first 64 states
(solutions of Eq.~(\ref{Eig})) are shown in Fig.~1 jointly with the
bare energies of the pigments. It is evident from this figure that
the lowest state is separated from the upper states by energy gap of
approximately 13 meV.

To determine the energy transfer efficiency, the spectral functions
of the environment should be defined. We use the Drude-Gaussian form
of the spectral function for the heat bath modes, as
\begin{equation}
J^{\rm ph}(\omega ) = \lambda^{\rm
ph}\left(\frac{\omega}{\omega_c}\right)\exp\left(
-\frac{\omega}{\omega_c}\right) .
\end{equation}
This type of spectral function is used both for the description of
the environment in photosynthetic complexes \cite{Fleming_Cho} and
for general condensed matter problems \cite{Leggett}. The spectral
function for the specific vibronic mode is given by
\begin{equation}
J^{\rm vib}(\omega ) = \lambda^{\rm vib}\Omega \delta ( \omega -
\Omega ) .
\end{equation}
The total spectral function, including the contributions of the heat
bath modes and a specific vibronic mode, is shown in Fig.~2. The
cutting frequency $\omega_c$ is taken to be $18.6$ meV and the
frequency of the vibronic mode is chosen to be $10$ meV for this
figure.

The efficiency, Eq.~(\ref{Eff}), is shown in Fig.~3 as a function of
the frequency of the vibronic mode. One can see that with decreasing
frequency, the efficiency starts to drop at approximately 16 meV.
This drop can be quite significant, up to 4 times at low
frequencies. We also show in this figure the efficiency value for
case when there is no strong coupling to the specific vibronic mode
which equals to 0.8314 for our set of parameters. For our
calculations, we use: the temperature $T = 77$ K, the light-Bchl
coupling strength $F = 10^{-4}$ meV (with the light coupled to the
first pigment only), refraction index $n_{\rm refr} = 1.42$, the
heat bath reorganization energy $\lambda^{\rm ph} = 4.34$ meV, the
reorganization energy of the vibronic mode $\lambda^{\rm vib} = 18.6
$ meV, coupling to the reaction center $g_{nk} = 6.6~10^{-3}$ meV
(with only the third pigment coupled), and the coupling to the
quenching field $Q_m = 6.6~10^{-4}$ meV.

To understand the physical reasons for such significant drop of the
energy transfer efficiency, we calculate the time dependence of the
pigment populations with the results shown in Fig.~4(a-f). For
initial conditions, we use the situation when the first pigment
(coupled to the light) is populated and all other Bchls are not.
With no strong coupling to the vibronic mode (Fig.~4a), there are
well-known quantum oscillations of populations of the first and
second pigments (for times of about 0.5 ps) and by the time of about
3 ps the exciton energy is mostly transferred to the third pigment
coupled to the reaction center. It should be noted that the fifth,
sixth, and seventh pigments remain unpopulated all the time. Almost
the same picture can be seen for the high frequency of the vibronic
mode (Fig.~4f). With the frequency decreasing, the fifth, sixth, and
seventh pigments start to be populated and at $\Omega = 4$ meV all
seven Bchls are populated almost equally by 3 ps. It should be
emphasized that the energies of the Bchl 3 and Bchl 4, Bchl 4 and
Bchl 5, and Bchl 5 and Bchl 6, are all separated by 16-19 meV.
Accordingly, the strong coupling to the vibronic modes of
appropriate frequency with the preservation of ``vibronic
coherence'' \cite{Mancal1,Mancal2,Plenio2} can lead to the formation
of ``polaronic modes'' between corresponding pigments. As the
coupling to the vibronic mode is strong, multi-phonon interaction is
possible. Formally, it corresponds to non-vanishing contributions of
the high-order Bessel functions of Eqs. (A15) and (A16). Moreover,
one can see from these equations that the interference between the
specific mode and heat bath modes is possible, therefore the energy
mismatch preventing the polaronic mode formation can be compensated
by the heat bath. As a result, the exciton, which is transferred to
Bchl 3 from Bchl 2, does not proceed to the reaction center, but
oscillates between Bchls 3, 4 5, and 6 and the energy is eventually
lost to the quenching field or the heat bath.

Of special interest is the suppression of the quantum beating
between Bchl 1 and Bchl 2 at $\Omega$ = 25 meV, Fig.~4e. It can be a
result of similar polaronic mode formation. The separation between
the energies of Bchl 1 and Bchl 6 is about 24 meV, so
vibronic-assisted transfer between these pigments is possible.
Correspondingly, the exciton proceeds to the third pigment not via
Bchl 2 but via Bchls 6, 5, and 4. It should be noted that the energy
efficiency is {\it not} suppressed at this frequency.

Our approach allows us to examine the situation when the vibronic
mode is deliberately coupled to a specific pigment. In Fig.~5, we
show the energy transfer efficiency as a function of the vibronic
mode frequency, when only the Bchl 3 is coupled to the vibronic
mode. One can see that the efficiency is suppressed at low
frequencies but not as drastically as in Fig.~3. The reason for such
relatively small suppression is that the polaronic mode is formed
between the Bchl 3 and Bchl 4 only, with no exciton transfer farther
to Bchls 5 and 6. However, it is evident from Fig.~5 that at
$\Omega$ = 30 meV, the energy transfer efficiency even exceeds the
value for no coupling to the vibronic mode. The corresponding time
dependencies of the pigment populations are shown in Fig.~6. For
$\Omega$ = 4 meV, Fig.~6a, Bchls 5 and 6 remain unpopulated, in
contrast to Fig.~4b, when all the sites are coupled to the vibronic
mode. One can see from Fig.~6b that for $\Omega$ = 30 meV the
population of Bchl 3 at 3 ps is even higher than that of the
unperturbed system, Fig.~4a.

\section{Conclusions}
In conclusion, we have developed an approach allowing to study the
exciton transfer through a network for the case when the
reorganization energy is of the order of the inter-site couplings.
Our method is not restricted to the one-particle case, so we can
describe the propagation of several excitons through the system. We
have taken into account the effects of radiative and quenching
baths, as well as the coupling to the reaction center,
perturbatively with the secular approximation. We have gone beyond
this approximation for the protein environment examining the
non-Markovian effects as well. While the majority vibronic modes
have been treated as a heat bath, we have also included the strong
coupling to a specific vibronic mode into consideration. For the
heat bath modes, we have used a high temperature (or low frequency)
approximation for the dynamics inside the non-Markovian integral,
while the specific mode was treated exactly. Accordingly, we have
determined the contributions of multi-phonon processes and obtained
that the contributions of the heat bath modes and the vibronic mode
are interconnected, as the frequency and the reorganization energy
of the vibronic mode is involved in the relaxation matrix for the
heat bath and vice versa.

We have obtained the efficiency of the energy transfer directly, as
the ratio of the energy transferred to the reaction center and the
total energy absorbed by the system. We have calculated this
efficiency for a specific system, the FMO complex, and found that
this efficiency is dropped significantly, if the energy of the
strongly-coupled vibronic mode becomes smaller than 16 meV. We have
attributed that to the formation of ``polaronic modes'' where the
exciton is transferred back and forth between two pigments with
absorption and emission of vibronic quanta. Accordingly, the exciton
instead of proceeding to the reaction center from the lowest-energy
Bchl 3, is transferred sequentially to Bchls 4, 5, and 6, and the
energy is eventually lost to the quenching field or to the
environment heat bath. We have illustrated this effect determining
the time dependencies of the pigment populations and showing that
for the low frequency of the strongly-coupled vibronic mode, the
exciton is almost equally distributed between all the pigments. We
have obtained the well-known oscillations in the populations of Bchl
1 and Bchl 2 and showed that these oscillations are suppressed at
the vibronic mode frequency of 30 meV. This corresponds to the
separation of the energies of Bchls 1 and 6 and we attribute this
effect to the polaronic mode between these pigments and the energy
transfer avoiding Bchl 2. It is interesting that it does not affect
the energy transfer efficiency. We have also shown that the
efficiency can even be enhanced if the specific vibronic mode is
coupled deliberately to Bchl 3. In this case, the obtained value
exceeds that for the unperturbed complex.

\appendix

\section{Contributions of various mechanisms to the evolution of the
density matrix}

In the Appendix, we provide the calculations of various
contributions to the equation of the motion for the density matrix,
Eq.~(\ref{em}). The chromophore sites are weakly coupled to the
radiative and quenching bath, as well to the reaction center. Hence,
the contribution of these components of the variable ${\cal A}$ to
the Eq.~(\ref{em}) can be treated perturbatively with the secular
approximation. Thus, for the diagonal part, $\rho_{\mu} =
\rho_{\mu\mu} = |\mu\rangle \langle \mu|,$ of the operator
$\rho_{\mu\nu}$ we obtain
\begin{eqnarray}
- i \langle [\rho_{\mu}, H_{{\rm rec}} + H_{{\rm trap}}]_- \rangle=
- \sum_{\alpha} ( \gamma_{\alpha \mu}^{{\rm rec}} + \gamma_{\alpha
\mu}^{{\rm trap}} )\langle \rho_{\mu}\rangle + \sum_{\alpha} (
\gamma_{\mu \alpha }^{{\rm rec}} + \gamma_{\mu \alpha }^{{\rm trap}}
)\langle \rho_{\alpha}\rangle.
\end{eqnarray}
Recombination events and the trapping of excitations by the reaction
center provide the following contribution to the dephasing of
excitonic degrees of freedom ($\mu\neq \nu$)
\begin{eqnarray}
- i \langle [\rho_{\mu\nu}, H_{{\rm rec}} + H_{{\rm trap}}]_-\rangle
= - ( \Gamma_{\mu \nu }^{{\rm rec}} + \Gamma_{\mu \nu }^{{\rm trap}}
)\langle \rho_{\mu\nu}\rangle.
\end{eqnarray}
The relaxation rates are given by
\begin{eqnarray}
\gamma_{\mu\alpha}^{{\rm rec}} = 2 \sum_n |a_{n;\alpha \mu}|^2 \chi_n''(\omega_0 + \omega_{\mu\alpha} ) n(\omega_0 + \omega_{\mu\alpha} ) + \nonumber\\
2 \sum_n |a_{n;\mu\alpha}|^2 \chi_n''(\omega_0 - \omega_{\mu\alpha}
) [ 1 + n(\omega_0 - \omega_{\mu\alpha} )]
\end{eqnarray}
and
\begin{eqnarray}
\gamma_{\mu\alpha}^{{\rm trap}} = \sum_n |a_{n;\alpha \mu}|^2
\Gamma_n^{{\rm trap}}(\omega_0 + \omega_{\mu\alpha})  n(\omega_0 +
\omega_{\mu\alpha} ) +
\nonumber\\
\sum_n |a_{n;\mu \alpha }|^2 \Gamma_n^{{\rm trap}}(\omega_0 -
\omega_{\mu\alpha}) [1 +  n(\omega_0 - \omega_{\mu\alpha} )].
\end{eqnarray}
The dephasing rates consist of two parts, $\Gamma_{\mu\nu}^{{\rm
rec}} = \Gamma_{\mu}^{{\rm rec}} + \Gamma_{\nu}^{{\rm rec}}, $ and
$\Gamma_{\mu\nu}^{{\rm trap}} = \Gamma_{\mu}^{{\rm trap}} +
\Gamma_{\nu}^{{\rm trap}}, $ where
\begin{eqnarray}
\Gamma_{\mu}^{{\rm rec}} = \sum_{n\alpha} |a_{n;\mu\alpha}|^2 \chi''_n(\omega_0 - \omega_{\mu\alpha} ) n(\omega_0 - \omega_{\mu\alpha}) + \nonumber\\
\sum_{n\alpha} |a_{n;\alpha\mu}|^2 \chi''_n(\omega_0 +
\omega_{\mu\alpha}) [ 1 +  n(\omega_0 + \omega_{\mu\alpha})],
\end{eqnarray}
and
\begin{eqnarray}
\Gamma_{\mu}^{{\rm trap}} = (1/2)\sum_{n\alpha} |a_{n;\mu\alpha}|^2 \Gamma_n^{{\rm trap}}(\omega_0 - \omega_{\mu\alpha}) n(\omega_0 - \omega_{\mu\alpha}) + \nonumber\\
(1/2)\sum_{n\alpha} |a_{n;\alpha\mu}|^2 \Gamma_n^{{\rm
trap}}(\omega_0 + \omega_{\mu\alpha}) [ 1 +  n(\omega_0 +
\omega_{\mu\alpha})].
\end{eqnarray}
In this expressions, $n(\omega ) = [\exp(\omega/T) - 1]^{-1}$ is the
Bose distribution, $\Gamma_n^{{\rm trap}}$  is the trapping rate
defined as
\begin{equation}
\Gamma_n^{{\rm trap}} = 2\pi \sum_k  |g_{kn}|^2 \delta (\omega -
\epsilon_k),
\end{equation}
and the imaginary part of the bath susceptibility,
$\chi_n''(\omega)$, contains contributions of the blackbody heat
bath and the Ohmic quenching bath, as
\begin{equation}
\chi_n''(\omega) = (2/3) n_{{\rm refr}} |d_n|^2 (\omega/c)^3 +
\alpha_n \omega,
\end{equation}
with $n_{{\rm refr}}$ being the refractive index of the medium.

The interaction with the protein environment cannot be considered
weak, and, correspondingly, we cannot use the secular approximation
employed in the previous subsection. Following
Ref.~\cite{SmJCP2011}, we introduce various spectral densities and
reorganization energies as
\begin{eqnarray}
J^{{\rm ph}}_{\mu}(\omega) = \sum_j \frac{m_j\omega_j^3}{2} \left( \sum_m C_{jm} \langle \mu|N_m |\mu\rangle \right)^2 \delta(\omega - \omega_j), \nonumber\\
\bar{J}^{{\rm ph}}_{\mu\nu}(\omega) = \sum_j \frac{m_j\omega_j^3}{2}
\left( \sum_m C_{jm} \langle \mu|N_m |\mu\rangle - \sum_m C_{jm}
\langle
\nu|N_m |\nu\rangle\right)^2 \delta(\omega - \omega_j) \nonumber\\
\tilde{J}^{{\rm ph}}_{\mu\nu}(\omega) = \sum_j
\frac{m_j\omega_j^3}{2} \left\vert\sum_m C_{jm} \langle \mu|N_m
|\nu\rangle\right\vert^2 \delta(\omega - \omega_j), \mu \neq \nu ,
\end{eqnarray}
\begin{eqnarray}
\lambda^{{\rm ph}}_{\mu} = \int_0^{\infty}\frac{d\omega}{\omega}
J^{{\rm ph}}_{\mu}(\omega)
= \sum_j \frac{m_j\omega_j^2}{2} \left( \sum_m C_{jm} \langle \mu|N_m |\mu\rangle \right)^2, \nonumber\\
\bar{\lambda}^{{\rm ph}}_{\mu\nu} =
\int_0^{\infty}\frac{d\omega}{\omega} \bar{J}^{{\rm
ph}}_{\mu\nu}(\omega) = \sum_j \frac{m_j\omega_j^2}{2} \left( \sum_m
C_{jm} \langle \mu|N_m |\mu\rangle - \sum_m C_{jm} \langle \nu|N_m
|\nu\rangle\right)^2,
\end{eqnarray}
and
\begin{eqnarray}
\lambda^{{\rm vib}}_{\mu}
= \frac{M\Omega^2}{2} \left( \sum_m C_{m} \langle \mu|N_m |\mu\rangle \right)^2, \nonumber\\
\bar{\lambda}^{{\rm vib}}_{\mu\nu}  = \frac{M\Omega^2}{2} \left(
\sum_m C_{m} \langle \mu|N_m |\mu\rangle - \sum_m C_{m} \langle
\nu|N_m
|\nu\rangle\right)^2, \nonumber\\
\tilde{\lambda}^{{\rm vib}}_{\mu\nu}  = \frac{M\Omega^2}{2}
\left\vert \sum_m C_{m} \langle \mu|N_m |\nu\rangle \right\vert^2,
\mu \neq \nu .
\end{eqnarray}

It was shown in Ref.~\cite{SmJCP2011} that the contribution of
diagonal environment fluctuations can be determined exactly and they
affect the off-diagonal elements of the density matrix only. With
inclusion of an additional vibronic mode, the time evolution caused
by these diagonal fluctuations has the form
\begin{equation}
\rho_{\mu\nu}(t) = \exp \left[ i\omega_{\mu\nu}\,t\right] \exp
\left[ -(\bar{\lambda}^{{\rm ph}}_{\mu\nu} + \bar{\lambda}^{{\rm
vib}}_{\mu\nu})\,T\,t^2\right] \rho_{\mu\nu}(0).
\end{equation}

We evaluate the internal dynamics in the non-Markovian integrals and
obtain the following contributions of non-diagonal environment and
vibronic fluctuations to Eq.~(\ref{em}). (It should be noted that
high temperature and low environment frequencies approximations of
Ref.~\cite{SmJCP2011} have been applied to the heat bath
contribution, not to that of the specific single mode.) The
evolution of the diagonal matrix elements is given by
\begin{eqnarray}
\langle -i [ \rho_{\mu}, H_{{\rm e-ph}}+H_{{\rm e-vib}}]_{-}\rangle
= - \sum_{\alpha} \left( \gamma^{{\rm ph}}_{\alpha\mu} +
\gamma^{{\rm vib}}_{\alpha\mu}\right) \langle \rho_{\mu} \rangle +
\sum_{\alpha} \left( \gamma^{{\rm ph}}_{\mu\alpha} + \gamma^{{\rm
vib}}_{\mu\alpha}\right) \langle \rho_{\alpha} \rangle.
\end{eqnarray}
For the off-diagonal elements, we obtain
\begin{eqnarray}
\langle -i [ \rho_{\mu\nu}, H_{{\rm e-ph}}+H_{{\rm
e-vib}}]_{-}\rangle = - (\Gamma^{{\rm ph}}_{\mu\nu}+\Gamma^{{\rm
vib}}_{\mu\nu}) \langle \rho_{\mu\nu}\rangle.
\end{eqnarray}
The relaxation matrices are given by
\begin{eqnarray}
\gamma^{{\rm ph}}_{\mu\alpha} = \sqrt{\frac{\pi}{\bar{\lambda}^{{\rm
ph}}_{\alpha\mu}T}}\exp \left[-\frac{\bar{\lambda}^{{\rm
vib}}_{\mu\alpha}}{\Omega}\coth \frac{\Omega}{2T} \right]
\sum_{l=-\infty}^{\infty} J_l\left[ \frac{\bar{\lambda}^{{\rm
vib}}_{\mu\alpha}}{\Omega}\right]
\int_0^{\infty} d\omega \,\tilde{J}^{{\rm ph}}_{\alpha\mu}(\omega)\,  n(\omega) \nonumber\\
\times \left( I_0\left[\frac{\bar{\lambda}^{{\rm
vib}}_{\mu\alpha}}{\Omega}\coth \frac{\Omega}{2T}\right]  \left(
\exp\left[ - \frac{(\omega + \Omega_{\alpha \mu} + l\Omega -
\bar{\lambda}^{{\rm ph}}_{\alpha\mu})^2}{ 4 \bar{\lambda}^{{\rm
ph}}_{\alpha\mu} T} \right] \right. \right. \nonumber
\\ + \left. \exp\left( \frac{\omega}{T}\right) \exp\left[ - \frac{(\omega -
\Omega_{\alpha \mu} - l\Omega + \bar{\lambda}^{{\rm
ph}}_{\alpha\mu})^2}{ 4 \bar{\lambda}^{{\rm ph}}_{\alpha\mu} T}
\right] \right)
\nonumber\\
+ \sum_{s=1}^{\infty} I_s\left[\frac{\bar{\lambda}^{{\rm
vib}}_{\mu\alpha}}{\Omega}\coth \frac{\Omega}{2T}\right] \left\{
\exp\left[ - \frac{(\omega + \Omega_{\alpha \mu} + (l+s)\Omega -
\bar{\lambda}^{{\rm ph}}_{\alpha\mu})^2}{ 4 \bar{\lambda}^{{\rm
ph}}_{\alpha\mu}
T} \right] \right. \nonumber \\
+ \exp\left[ - \frac{(\omega + \Omega_{\alpha \mu} + (l-s)\Omega -
\bar{\lambda}^{{\rm ph}}_{\alpha\mu})^2}{ 4 \bar{\lambda}^{{\rm
ph}}_{\alpha\mu}
T} \right] \nonumber \\
+  \exp\left( \frac{\omega}{T}\right) \left( \exp\left[ -
\frac{(\omega + \Omega_{\alpha \mu} + (l+s)\Omega -
\bar{\lambda}^{{\rm ph}}_{\alpha\mu})^2}{ 4 \bar{\lambda}^{{\rm
ph}}_{\alpha\mu} T} \right] \right. \nonumber \\ + \left. \left.
\left. \exp\left[ - \frac{(\omega + \Omega_{\alpha \mu} +
(l-s)\Omega - \bar{\lambda}^{{\rm ph}}_{\alpha\mu})^2}{ 4
\bar{\lambda}^{{\rm ph}}_{\alpha\mu} T} \right] \right) \right\}
\right)
\nonumber \\
\end{eqnarray}
and
\begin{eqnarray}
\gamma^{{\rm vib}}_{\mu\alpha} =
\sqrt{\frac{\pi}{\bar{\lambda}^{{\rm ph}}_{\alpha\mu}T}}\exp
\left[-\frac{\bar{\lambda}^{{\rm vib}}_{\mu\alpha}}{\Omega}\coth
\frac{\Omega}{2T} \right] \sum_{l=-\infty}^{\infty} J_l\left[
\frac{\bar{\lambda}^{{\rm vib}}_{\mu\alpha}}{\Omega}\right]
\Omega \, \tilde{\lambda}^{{\rm vib}}_{\alpha\mu}\,  n(\Omega) \nonumber\\
\times \left( I_0\left[\frac{\bar{\lambda}^{{\rm
vib}}_{\mu\alpha}}{\Omega}\coth \frac{\Omega}{2T}\right]  \left(
\exp\left[ - \frac{(\Omega + \Omega_{\alpha \mu} + l\Omega -
\bar{\lambda}^{{\rm ph}}_{\alpha\mu})^2}{ 4 \bar{\lambda}^{{\rm
ph}}_{\alpha\mu} T} \right] \right. \right. \nonumber
\\ + \left. \exp\left(
\frac{\Omega}{T}\right) \exp\left[ - \frac{(\Omega - \Omega_{\alpha
\mu} - l\Omega + \bar{\lambda}^{{\rm ph}}_{\alpha\mu})^2}{ 4
\bar{\lambda}^{{\rm ph}}_{\alpha\mu} T} \right] \right)
\nonumber\\
+ \sum_{s=1}^{\infty} I_s\left[\frac{\bar{\lambda}^{{\rm
vib}}_{\mu\alpha}}{\Omega}\coth \frac{\Omega}{2T}\right] \left\{
\exp\left[ - \frac{(\Omega + \Omega_{\alpha \mu} + (l+s)\Omega -
\bar{\lambda}^{{\rm ph}}_{\alpha\mu})^2}{ 4 \bar{\lambda}^{{\rm
ph}}_{\alpha\mu}
T} \right] \right. \nonumber \\
+ \exp\left[ - \frac{(\Omega + \Omega_{\alpha \mu} + (l-s)\Omega -
\bar{\lambda}^{{\rm ph}}_{\alpha\mu})^2}{ 4 \bar{\lambda}^{{\rm
ph}}_{\alpha\mu}
T} \right] \nonumber \\
+ \exp\left( \frac{\Omega}{T}\right) \left( \exp\left[ -
\frac{(\Omega + \Omega_{\alpha \mu} + (l+s)\Omega -
\bar{\lambda}^{{\rm ph}}_{\alpha\mu})^2}{ 4
\bar{\lambda}^{{\rm ph}}_{\alpha\mu} T} \right] \right. \nonumber \\
\left. \left. \left. + \exp\left[ - \frac{(\Omega + \Omega_{\alpha
\mu} + (l-s)\Omega - \bar{\lambda}^{{\rm ph}}_{\alpha\mu})^2}{ 4
\bar{\lambda}^{{\rm ph}}_{\alpha\mu} T} \right] \right) \right\}
\right),
\nonumber \\
\end{eqnarray}
where
\begin{equation}
\Omega_{\alpha \mu} = \omega_{\alpha \mu} + \bar{\lambda}^{{\rm
ph}}_{\mu} - \bar{\lambda}^{{\rm ph}}_{\alpha} + \bar{\lambda}^{{\rm
vib}}_{\mu} - \bar{\lambda}^{{\rm vib}}_{\alpha}
\end{equation}
and $J_l(z)$ and $I_s(z)$ are the ordinary and modified Bessel
functions, respectively. It should be emphasized that the heat bath
and vibronic contributions are interconnected, as the vibronic mode
is involved in the expression for $\gamma^{{\rm ph}}_{\mu\alpha}$
and vise versa.

Similar to Ref.~\cite{SmJCP2011}, the dephasing rates can be
expressed as
\begin{eqnarray}
\Gamma^{{\rm ph,vib}}_{\mu\nu} = \Gamma^{{\rm ph,vib}}_{\mu} +
\Gamma^{{\rm ph,vib}}_{\nu}, \nonumber \\
\Gamma^{{\rm ph,vib}}_{\mu} = \frac{1}{2} \sum_{\alpha} \gamma^{{\rm
ph,vib}}_{\alpha\mu}.
\end{eqnarray}

\vspace{2cm}

\noindent \textbf{Acknowledgements}

This work is partially supported by the RIKEN iTHES Project, MURI
Center for Dynamic Magneto-Optics, JSPS-RFBR contract no.
12-02-92100, Grant-in-Aid for Scientific Research (S), MEXT Kakenhi
on Quantum Cybernetics and the JSPS via its FIRST program. L. M. was
also partially supported by PSC-CUNY award 65245-00 43.

\pagebreak

\begin{figure}
\includegraphics[width=17cm]{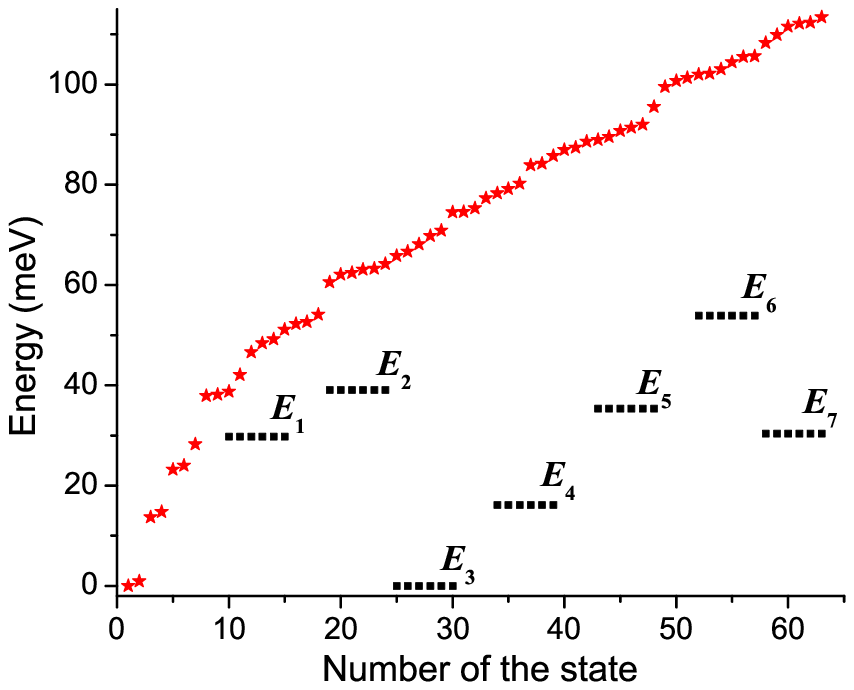}
\vspace*{-2cm} \caption{(Color online) Eigenenergies (red stars) of
the unperturbed Hamiltonian. The bare energies of the seven pigments
of the FMO complex are shown as horizontal black dashed segments.}
\end{figure}
\begin{figure}
\includegraphics[width=17cm]{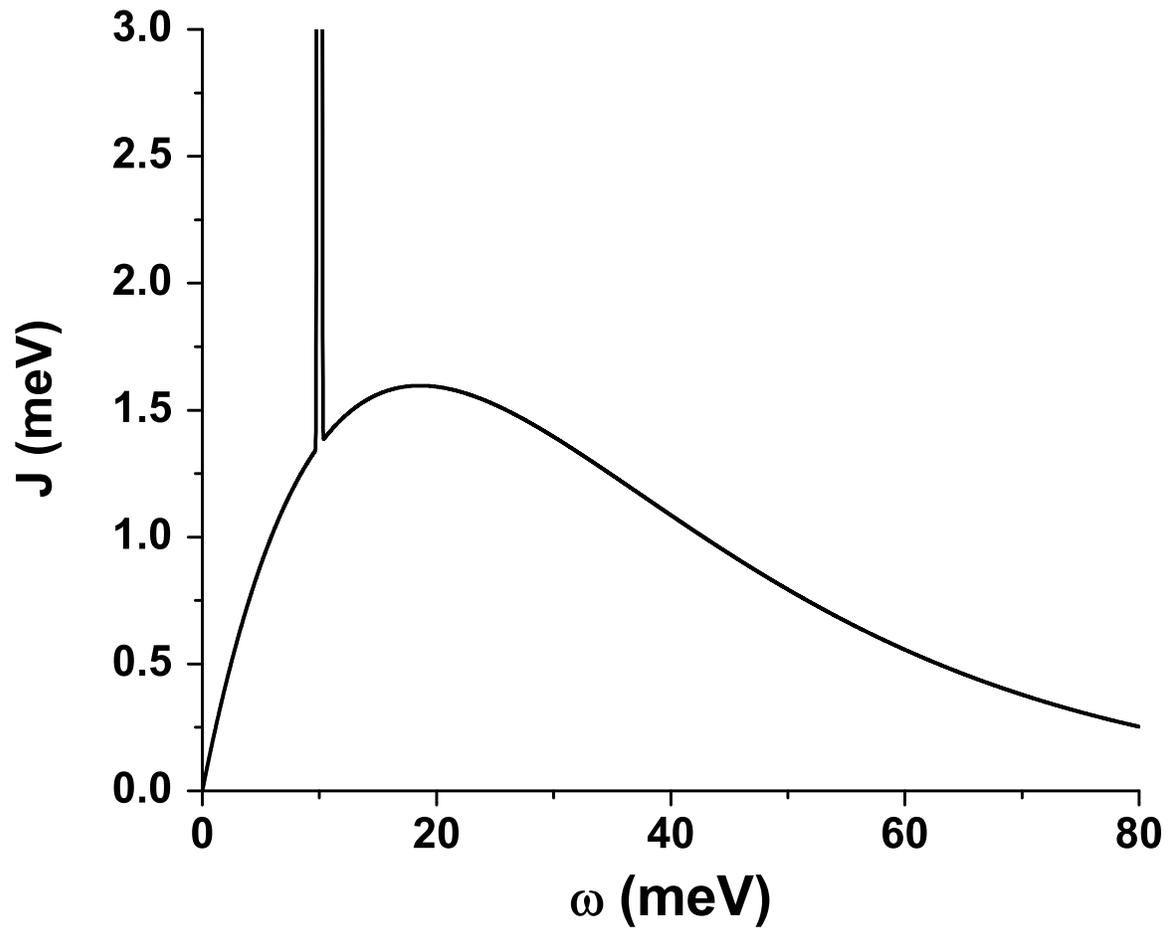}
\vspace*{-2cm} \caption{(Color online) Environment spectral function
$J(\omega)$ including contributions from the heat bath and a
specific vibronic mode with frequency $\Omega$ = 10 meV.}
\end{figure}
\begin{figure}
\includegraphics[width=17cm]{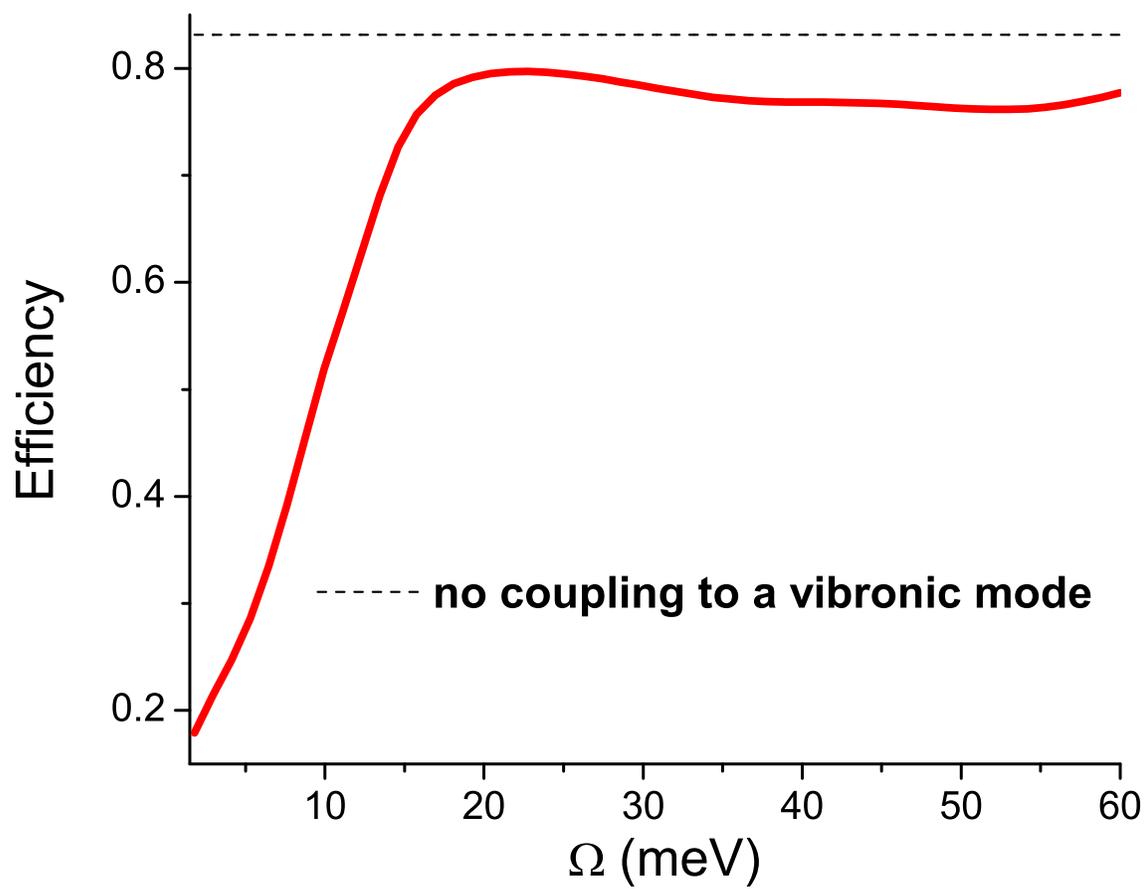}
\vspace*{-2cm} \caption{(Color online) Energy transfer efficiency as
a function of the vibronic mode frequency $\Omega$.}
\end{figure}
\begin{figure}
\includegraphics[width=8cm]{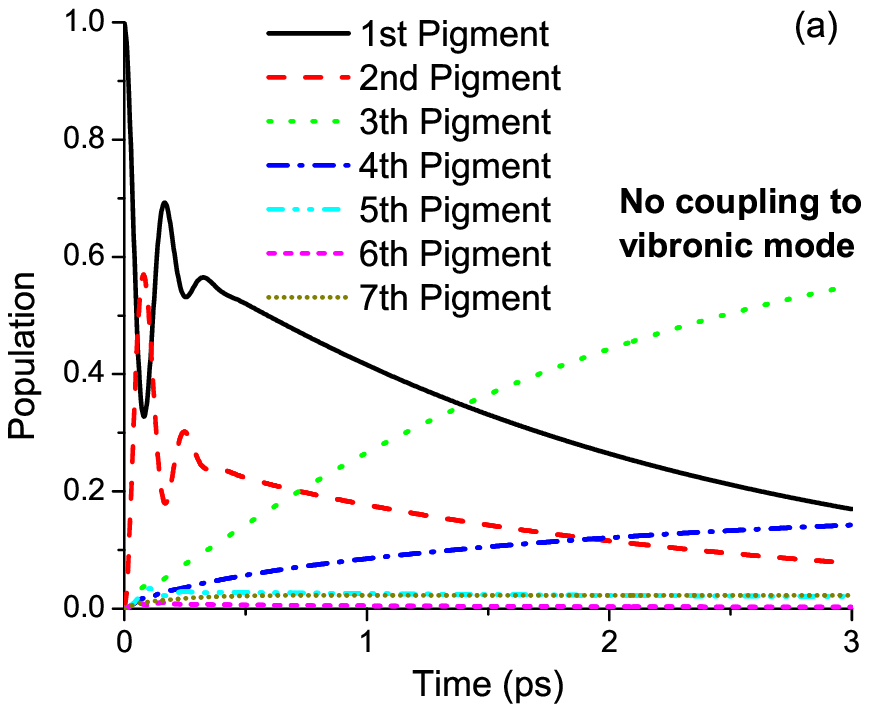}\hspace*{0.5cm}\includegraphics[width=8cm]{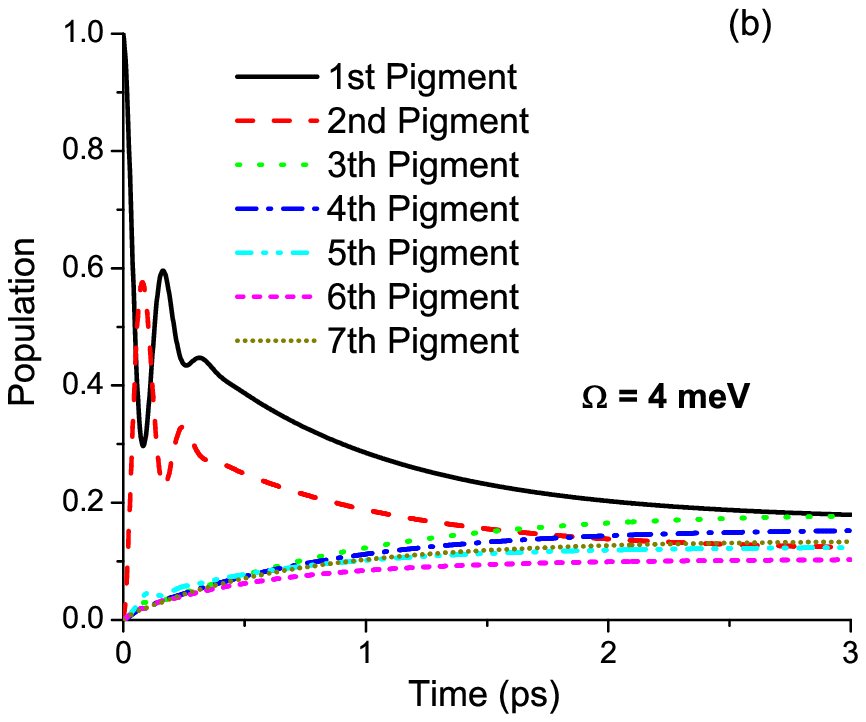}
\vspace*{0.5cm}
\includegraphics[width=8cm]{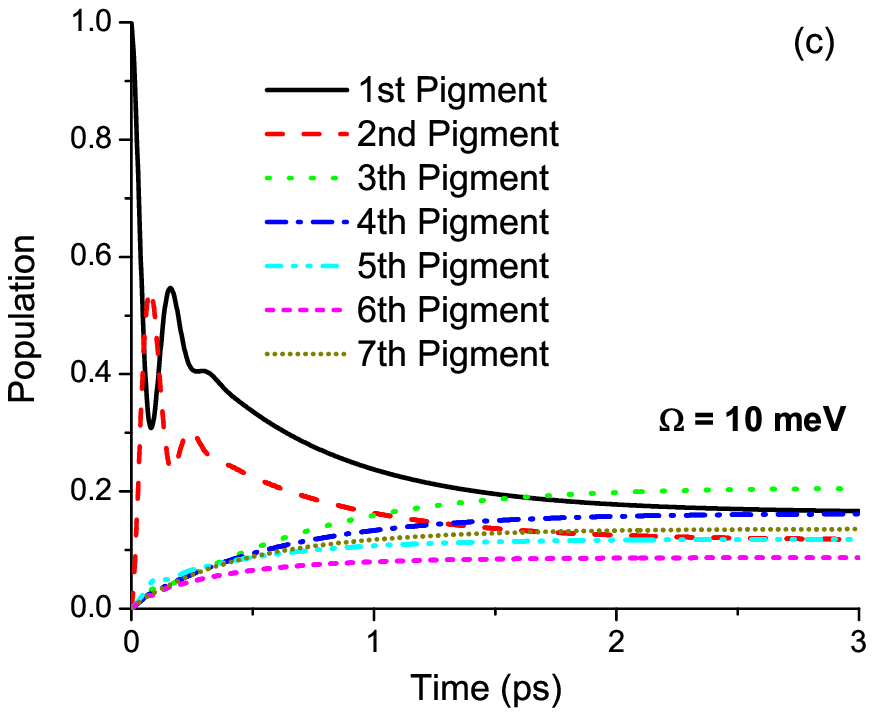}\hspace*{0.5cm}\includegraphics[width=8cm]{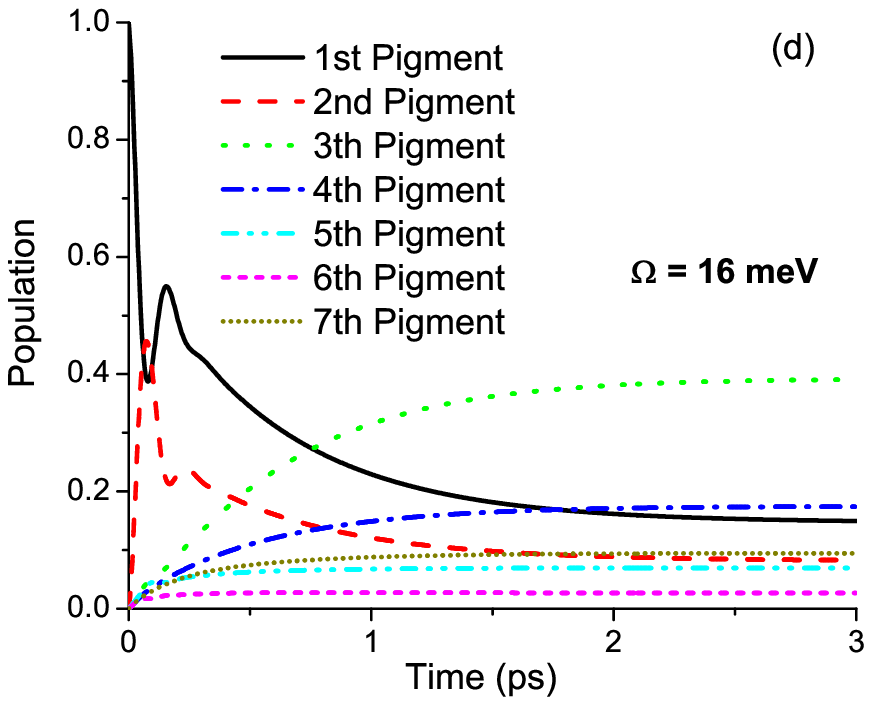}
\vspace*{0.5cm}
\includegraphics[width=8cm]{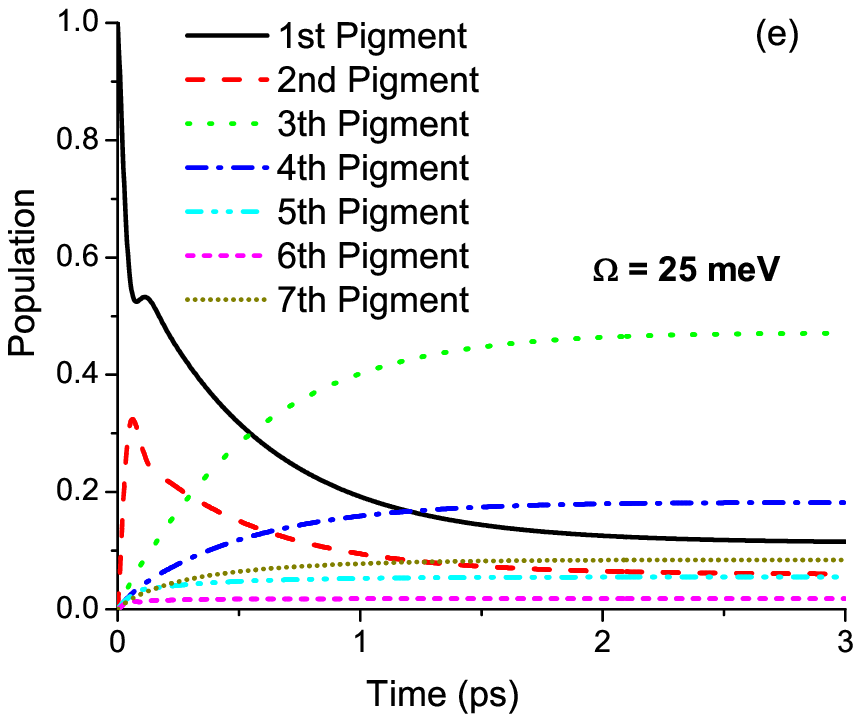}\hspace*{0.5cm}\includegraphics[width=8cm]{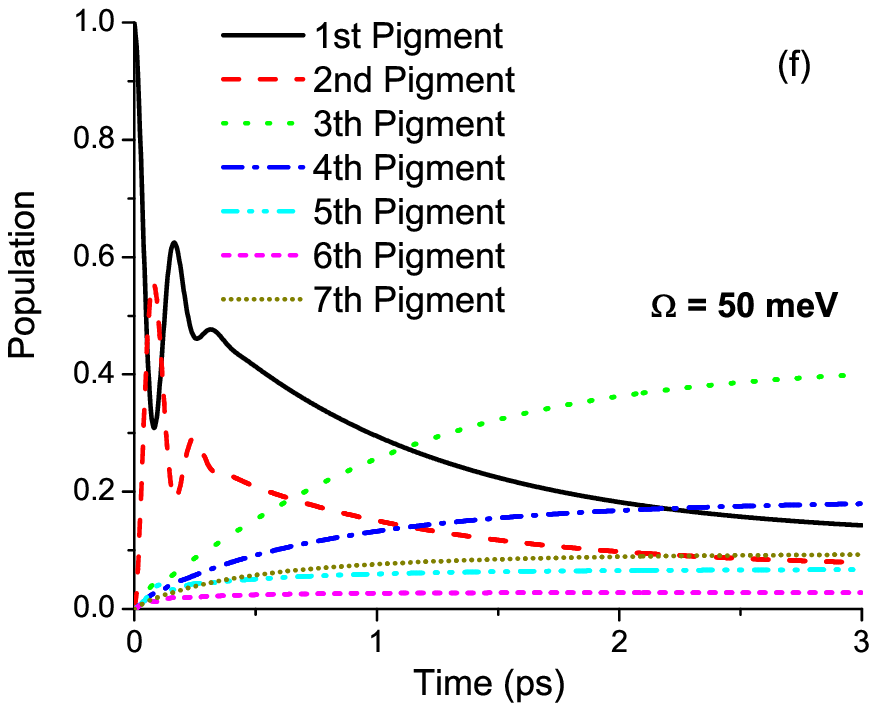}
\vspace*{-2cm} \caption{(Color online) Time dependencies of the
pigment populations for various frequencies of the vibronic
mode.}EPS
\end{figure}
\begin{figure}
\includegraphics[width=17cm]{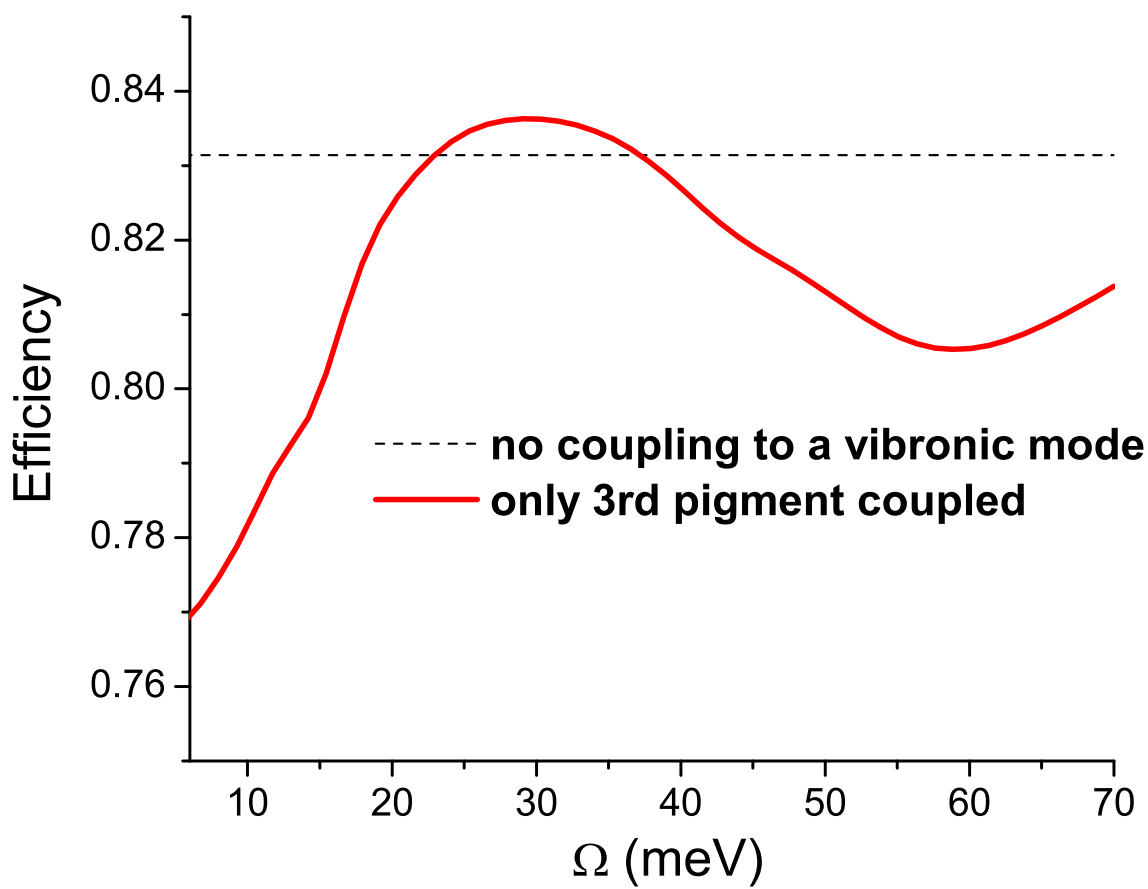}
\vspace*{-2cm} \caption{(Color online) Energy transfer efficiency as
a function of the frequency $\Omega$ of the vibronic mode coupled
only to the third pigment.}
\end{figure}
\begin{figure}
\includegraphics[width=8cm]{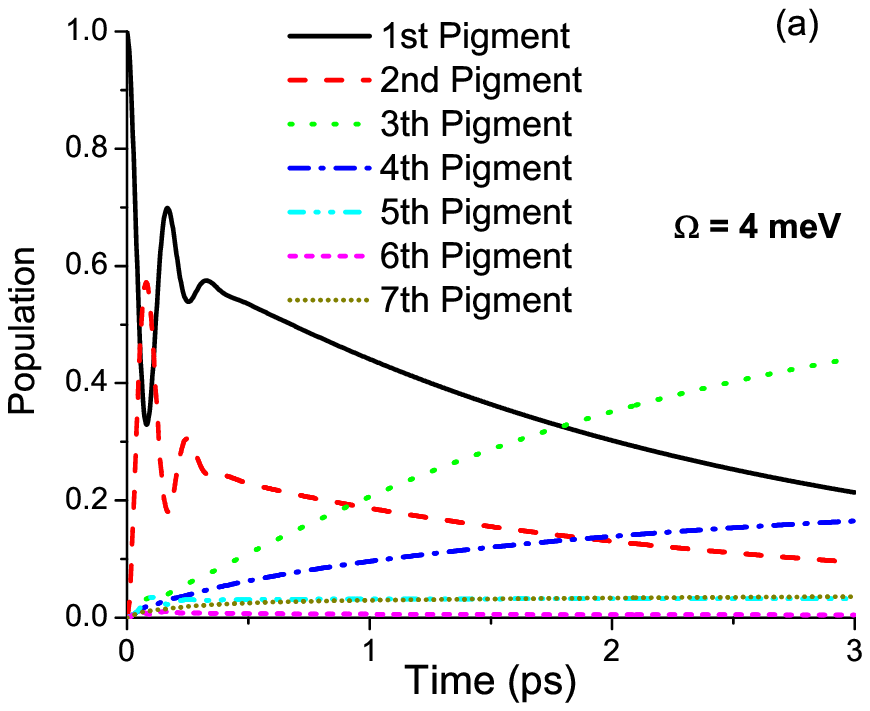}\hspace*{0.5cm}\includegraphics[width=8cm]{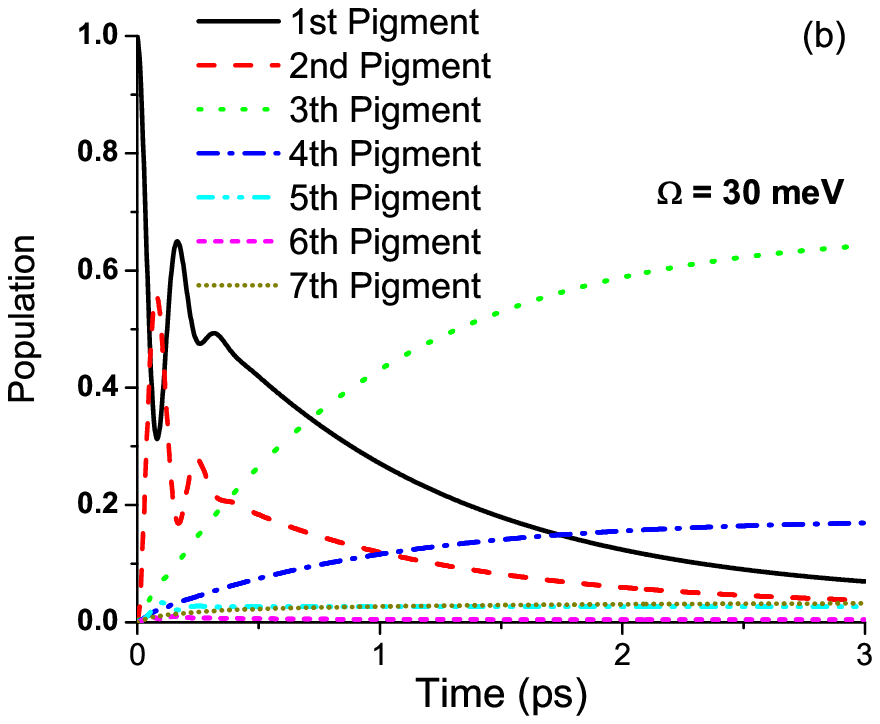}
\vspace*{-2cm} \caption{(Color online) Time dependencies of the
pigment populations for various frequencies of the vibronic mode
coupled only to the third pigment.}
\end{figure}

\end{document}